\documentclass[12pt]{iopart}

\usepackage{pictex}
\usepackage{iopams}

\newcommand{\be}{\begin{equation}}
\newcommand{\ee}{\end{equation}}
\newcommand{\tbar}{\overline{T}}
\newcommand{\h}{\bar{h}}

\newcommand{\salv}{\emph{Salviati: }}
\newcommand{\simp}{\emph{Simplicio: }}
\newcommand{\sagr}{\emph{Sagredo: }}

\begin{document}
version of \today

 \title[Surface Vacuum Energy in Cutoff Models]
 {Surface Vacuum Energy in Cutoff Models:\\  
Pressure Anomaly and\\
 Distributional Gravitational Limit}

 \author{Ricardo Estrada$^1$, Stephen A Fulling$^{2, 3}$
and    Fernando D Mera$^2$}

 \address{$^1$ Department of Mathematics, Louisiana State University,
     Baton Rouge, LA, 70803-4918 USA}

 \address{$^2$ Department of Mathematics, Texas A\&M University,
 College Station, TX, 77843-3368 USA} 
 
 \address{$^3$ Department of Physics, Texas A\&M University,
 College Station, TX, 77843-4242 USA} 

    \begin{abstract}
 Vacuum-energy calculations with ideal reflecting 
boundaries are plagued by boundary divergences, which presumably 
correspond to real (but finite) physical effects occurring near the 
boundary.   Our working hypothesis is that the stress tensor 
for idealized boundary conditions with some finite cutoff 
should be  a reasonable ad hoc model for the true situation.  
The theory will have a sensible renormalized limit when the cutoff 
is taken away; this requires making sense of the Einstein equation 
with a distributional source.
 Calculations with the standard ultraviolet cutoff reveal an 
inconsistency between energy and pressure similar to the one that 
arises in noncovariant regularizations of cosmological vacuum 
energy.
The problem disappears, however, if the cutoff is a spatial point 
separation in a ``neutral'' direction parallel to the boundary.
 Here we demonstrate these claims in detail, first for a single flat 
reflecting wall intersected by a test boundary, then more rigorously 
 for a region of finite cross section surrounded by four reflecting 
walls.
We also show how the moment-expansion theorem can be applied to the 
 distributional limits of the source and the solution of the 
 Einstein equation, resulting in a mathematically consistent differential 
equation where cutoff-dependent coefficients have been identified 
as renormalizations of properties of the boundary. 
 A number of issues surrounding the interpretation of these results 
are aired.
 \end{abstract}


 \maketitle

\section{Energy density and pressure near a plane boundary}
 \label{onewall}

\subsubsection*{Introductory remarks.} 
The effects of reflecting boundaries in quantum field theory
 continue to be of intense interest, from the mathematical to the 
experimental \cite{lamRL,milRL,Losalamos,qfe11}.
 
 From a mathematical point of view there are two kinds of effects.
 First, there is the classic Casimir energy, which is distributed 
throughout space,
associated semiclassically with periodic classical orbits,
 and finite after subtraction of the zero-point energy of infinite 
space. (Here we consider only space-time that is locally flat, 
and without any external potentials that would create additional 
distributed divergences.)
 Second, there is energy that is concentrated near the boundaries, 
associated with short paths, closed but not periodic, that bounce 
off the boundaries, and usually divergent in this sense:
 For a perfectly reflecting boundary (such as a perfect conductor in 
electromagnetism or a Dirichlet or Neumann boundary for a scalar 
field),
the vacuum expectation value of the energy density has a 
\emph{nonintegrable} behavior,
 \be 
  \rho \sim \frac{c_1}{s^4} + \frac{c_2}{s^3} + \cdots,
\label{bdryseries} \ee
 where $s$ is the distance from the boundary.
 (The leading exponent is $d+1$ when the spatial dimension is~$d$.)

  In the original Casimir scenario of the electromagnetic field 
between parallel flat conducting plates~\cite{cas,BrMa},
  energy of the second type is absent. In the famous Boyer 
  calculation \cite{boyer} for the electromagnetic field inside 
and 
  near a thin spherical shell of conductor, the total energy again 
  comes out finite, though only because of a sequence of special 
  cancellations \cite{DC}  (between electric and magnetic modes 
for 
  $c_1\,$, between interior and exterior for $c_2$ and  $c_4\,$,
and for a variety of complicated geometrical reasons 
\cite{systemat,BGH}
  for $c_3$).
 In general the divergent boundary contributions are present and 
are playing a major role in the ongoing overhaul of the theoretical 
ideas forced by the huge improvements in experimental results since 
1997.

 ``Analytic'' regularization methods (dimensional regularization 
and zeta functions) usually yield finite values for 
total energy, and ultraviolet-cutoff calculations can reach 
the same values by discarding the divergent leading terms in an 
expansion in powers of the cutoff parameter; 
but much current opinion tends to limit the scientific validity 
of such 
methods to calculations of the forces between disjoint, rigid 
bodies.
 In studying deformable bodies or gravitational effects, the actual 
energy density near the boundary must be taken seriously, together 
with the energy of the boundary material itself, from which it 
cannot be cleanly separated.
 It is agreed that divergences result from overidealization of the 
physics of the boundary.  For example, a real electromagnetic 
conductor is not perfectly conducting at arbitrarily high 
frequencies, and at truly small scales its atomic structure becomes 
relevant.
For a realistic boundary the effects represented crudely by inverse 
powers in \eref{bdryseries} are finite, but perhaps large.

In our present program we attempt to salvage some of the 
mathematical advantages of the old theory by hypothesizing 
that the vacuum stress tensor, $T^{\mu\nu}(x)$, of a quantum field 
interacting with an idealized boundary with an exponential 
ultraviolet or similar cutoff kept \emph{finite} is a reasonable 
\emph{ad hoc} model for a realistic physical situation. 
 We are particularly interested in the interpretation of the 
boundary divergences in the context of general relativity, an issue 
raised in the seminal paper of Deutsch and Candelas~\cite{DC}.
 Since the actual gravitational effects of vacuum energy in the 
laboratory are surely very small, the problem is primarly one of 
establishing consistency of the properly interpreted theory,
 so this approach should be sufficient --- certainly much more 
satisfactory than simply discarding divergent quantities that are 
integrals of perfectly finite and plausible local densities.
  The expectation is that the theory will have a sensible 
  renormalized limit when the cutoff is taken away;
 this requires making sense of the Einstein equation with a 
distributional source, interpreting the cutoff as a 
``regularization'' of distributions in the mathematical 
sense~\cite{EF2001}.
(See also the discussions of mass renormalization in
\cite{fallleip,fall2}.)

 This program was initiated in \cite{Leipzig}.  Detailed 
calculations of the stress tensor in a model with two spatial 
dimensions were presented in~\cite{rect}, where complete 
consistency was claimed between force calculations based on the 
pressure and those based on differentiation of total energy with 
respect to a geometrical parameter.
 However, close inspection later revealed a sign discrepancy for 
one pair of terms, those related to the leading term in the cutoff 
boundary energy near a boundary;
 in dimension 3 this problem persists, compounded by a discrepant 
factor of~$2$ \cite{safqfe09}.  
 There is no evidence that this anomaly affects the finite terms 
 constituting the standard Casimir force.  
 Nevertheless, since both methods of computing forces are widely 
used and assumed to be equivalent~\cite{mla}, understanding the 
phenomenon is imperative.
 The calculations in \cite{rect} and \cite{safqfe09} used a 
exponential ultraviolet cutoff, roughly equivalent to 
 ``point-splitting'' regularization in the time direction.
 It is reasonable to think that the trouble is related to the 
resulting violation of relativistic invariance.
 Therefore, a subprogram was launched to consider point-splitting 
in other directions~\cite{safqfe11}.
 For rectangular configurations it has been found that 
 a cutoff in a ``neutral'' direction (neither the time, nor the 
direction whose pressure is being sought) yields consistent and 
physically plausible results.  
 The purpose of the present article is to present detailed results 
along this line, together with recapitulation and updating of the 
material previously published only in conference proceedings 
\cite{Leipzig,safqfe09,safqfe11}.

 In this section we review the basic theory of 
 the vacuum stress tensor of a scalar field and its coupling to the 
gravitational field in linear approximation, and we 
 study the energy-pressure relations and  distributional limits 
associated to a plane Dirichlet boundary.
 The physical and philosophical issues raised are discussed in 
\sref{implications}.
The final section takes a more complete look at the energy-pressure 
relations in a rectangular box (more precisely, a waveguide) when 
the roles of all four boundaries are carefully taken into account.

\subsubsection*{Basic equations: quantum field theory.}  
Generalized to curved space-time, the action, Lagrangian function, 
and stress tensor of a scalar field in a cavity $\Omega$ are
\begin{equation}\fl
S= \int_{\Omega}  L\, \sqrt{ g } \,d^{d+1}x,
\qquad
L =-\,  \frac{1}{2}   [ g^{  \mu\nu } \partial_{\mu }   \phi 
  \partial_{\nu }  \phi
  + \xi R  \phi^2 ], \qquad \quad
  T^{\mu\nu} = \frac{2}{\sqrt{ g } } \,
  \frac{\delta S}{\delta g_{\mu\nu} } \,.   
\end{equation}
 We adopt the metric convention in which $g_{00}<0$.
The parameter $\xi$ labels different gravitational couplings. 
  In the flat-space 
limit the field equation and (classical) total energy are 
independent of~$\xi$, but the stress tensors are different.
 More specifically, that field equation is
\begin{equation}
 \frac{ \partial^2 \phi  }{\partial t^2}  =  \nabla^2  \phi 
 \end{equation}
accompanied by boundary conditions on $\Omega$, which we shall 
take to be Dirichlet here ($\phi=0$).
The simplest formulas for the stress tensor are obtained when 
$\xi=\frac14\,$,
 which corresponds to replacing the term $(\nabla\phi)^2$ in the 
energy density for minimal coupling ($\xi=0$) by $-
\phi\nabla^2\phi$, to which it is related by integration by parts:
  \begin{equation} 
T_{00}\left(\textstyle\frac14\right) = \frac{1}{2} \biggl [ \biggl  
( \frac{\partial \phi }{\partial  t }    \biggr )^2 - \phi \nabla^2 
\phi \biggr ], \ee \be T_{jj}\left(\textstyle\frac14\right) = 
\frac{ 1}{ 2}  
 \biggl [ \biggl ( \frac{ \partial \phi }{ \partial x_{j}  } 
 \biggr )^2 - \phi \frac{  \partial^2 \phi }{ \partial x_{j}^2 } 
\biggr ] \quad\mbox{for $j\ne0$}.
\end{equation}
 For other values of~$\xi$ one has
\begin{equation}
T_{\mu\nu}(\xi) = T_{\mu\nu } \left ( \textstyle\frac14  \right ) + 
\Delta T_{\mu \nu} \,  ,
\end{equation}
\begin{equation}
   \Delta T_{00} = -2 \left(\xi- \textstyle\frac14 \right ) 
\nabla \cdot ( \phi \nabla \phi ). 
\label{xirho}\end{equation}
\begin{equation}
\Delta T_{jj}=  - 2 \left(\xi- \textstyle\frac14 \right )
 \biggl [ \biggl ( \frac{ \partial \phi }{ \partial t } \biggr )^2 
- \sum_{k \neq j  } \biggl ( \frac{ \partial  \phi }
 { \partial x_{k } } \biggr )^2 
+ \phi \frac{  \partial^2 \phi }{ \partial x_{j}^2 } \biggr ] . 
\label{xip}\end{equation}                        
Henceforth we consider $\xi=\frac14$ unless otherwise indicated.  
The off-diagonal components of the tensor play no role in this paper.

The vacuum expectation values of components of the stress tensor 
can be expressed in terms of the Green function (called cylinder 
kernel or Poisson kernel)
\begin{equation}
 \overline{ T} ( t, \mathbf{ r} ,\mathbf{ r'} )  =
   - \sum_{n=1}^\infty \frac{1}{\omega_n}\,
\phi_n(\mathbf{ r} )\phi_n(\mathbf{ r' } )^* e^{-t\omega_n},
\label{tbar}\end{equation}
 where $\phi_n$ and $\omega_n$ are the eigenfunctions and 
eigenfrequencies of the cavity.
The formulas are
\be
   \rho=  \langle T_{00}\rangle = -\,\frac{1}{2} \, 
\frac{\partial^2 \overline{T}  }{\partial   t^2}\,, 
\label{rho}\ee
 \be  p_j=\langle T_{jj}\rangle = 
 \frac{1}{8} \left(\frac{\partial^2\overline{ T} }
 {\partial x_j{}\!^2} +
\frac{\partial^2 \overline{ T} }{\partial {x_j'}^2} -
2\,\frac{\partial^2 \overline{ T}  }{\partial x_j\,\partial 
x'_j}\right),
\label{pj} \ee
where it is understood, formally,  that $\mathbf{r'}$ is set equal to 
$\mathbf{r}$ and $t$ to $0$ at the end.
 Henceforth the angular brackets will usually  be omitted from 
expectation values.

 For more details of the foregoing formalism see~\cite{rect}.

In \eref{tbar}--\eref{pj}  $t$ is an ultraviolet cutoff parameter,
  not the physical time.  However, it can be thought of as  arising 
  by a Wick rotation from the difference of two physical time 
coordinates: $t = -i(x^0-x^{\prime0})$.
 In the standard ultraviolet-cutoff approach, $t$ is kept different 
from $0$ until the latest possible moment, so that 
 \eref{rho}--\eref{pj} define a nonsingular function even when
$\mathbf{r'}=\mathbf{r}$.
 On the other hand, if one thinks of 
 $x^0$ as just another coordinate in parallel to the components 
($x_j$) of $\mathbf{r}$, then it is natural to consider 
implementing a cutoff by separating the primed and unprimed points in 
some other direction in space-time.
At the spectral level this maneuver does not lead to a 
classically convergent summation like~\eref{tbar}, and if the 
 separation is in a spatial direction, this technical problem 
persists even if the
spatial coordinate in question is subjected to a Wick-like 
analytic continuation.  The series will converge in a 
distributional sense, however.  Moreover, 
 when a closed-form expression for $\tbar$ is available, as it is 
in all cases studied in this paper,
 the prescription can  be 
applied to it  without encountering any divergence.

 In its original application to a theory in curved space-time 
without boundaries~\cite{christensen},
the point separation could be in an arbitrary space-time  direction, 
 parametrized by a tangent vector~$tu^\mu$.
 In the particular case of flat space-time the resulting formula of 
Christensen is
\begin{equation}
 T_{\mu\nu}=\frac{1}{2\pi^2 t^4}  \biggl (  g_{\mu\nu} 
 - 4 \frac{  u_\mu u_\nu }{  u_{\rho }  u^{\rho} } \biggr )   .
 \end{equation}
The usual interpretation is that the direction-dependent term 
averages to~$0$, while the scalar cutoff parameter~$t$ is 
arbitrary;
thus  the stress tensor of flat space is
\be  T^{\mu\nu}_0 = \Lambda g^{\mu\nu},
 \label{lambdaterm}\ee
 where $\Lambda$ is a cosmological constant to be determined by 
observation.
(For an up-to-date discussion of this situation see~\cite{HJMM}.)
 Our present concern, however, is with a static boundary, and it 
seems that in this context the only sensible splitting directions 
are  pure time (real or imaginary) and all the spatial 
directions parallel to the boundary surface.\footnote{To 
foreshadow: From \eref{Mdef}--\eref{wallp}
  it can be seen that if the separation 
mixes space and real time, the denominators  will vanish somewhere in 
the  physical region, which defeats the purpose of a cutoff.
 Splitting involving $z-z'$ is ambiguous at best and 
problematical to define when the point is closer to the boundary 
than the cutoff distance.}

\subsubsection*{A reflecting wall intersected by a test wall.} 
 Consider a plane Dirichlet boundary at $z=0$.  We write 
 $\mathbf{ r}_{  \perp} =  (x,y) = (x_1,x_2)$ for the coordinates 
parallel to the wall.
The cylinder kernel for empty Minkowski space $\mathbb{R}^{3+1}$ is
 \be
  \overline{ T}_{0}   = -\,\frac{1}{2\pi^2}
\,\frac{1}{  t^2 + |\mathbf{ r}-\mathbf{ r'  } |^2}\,.
\label{Tfree}\ee
 The kernel for the half-space $z>0$ with the reflecting boundary 
is instantly obtained by the method of images, by adding to 
$\tbar_0$ the reflected term
 \be
 \overline{T} = \frac{1}{2\pi^2}\,\frac1{t^2 + | \mathbf{ r}_{\perp} 
- \mathbf{ r'}_{\perp}|^2 +(z+z')^2}\,.
\label{Twall}\ee
 For determining the physically relevant stress tensor one needs 
only the term \eref{Twall}, because the contribution of $\tbar_0$ 
is just \eref{lambdaterm}, ubiquitous and already considered.
 Therefore, we now reserve the notation $\tbar$ for the reflection 
term alone (sometimes called the ``renormalized'' cylinder kernel).

 We can now calculate the energy density and pressure from \eref{rho} 
and~\eref{pj}. After the differentiation, 
  without loss of generality  we  set $\mathbf{r}_\bot'=0$
 (because of translation invariance) and $z'=z$ (because we will 
not consider point-splitting perpendicular to the boundary).
 Then $t$ and the components of $\mathbf{r}_\bot$ are still 
available as cutoff parameters.
 We define
 \be M  = t^2+x^2+y^2 +4z^2.
 \label{Mdef}\ee
Then the results are
 \be
 2\pi^2\rho = M^{-3} [-3t^2 + x^2 + y^2 +4z^2], 
 \label{wallrho}\ee
\be\fl
  2\pi^2 p_1 =M^{-3} [-t^2+ 3x^2-y^2-4z^2],
 \quad
  2\pi^2 p_2 =M^{-3} [-t^2 -x^2 +3y^2-4z^2],
  \label{wallp}\ee
and $ p_3 = 0$.
Qualitatively, one can say that in the vacuum expectation there is 
a layer of energy against the wall, where $M$ is smallest.
 The vanishing of $p_3$ is understandable, because a
 rigid perpendicular displacement of the wall does not change the total 
energy of the configuration.

\begin{figure}
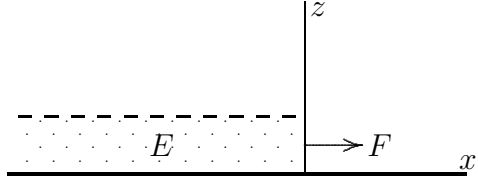

 \centerline{
 \beginpicture
 \setcoordinatesystem units <0.3in,0.3in> point at 1 0
 \setplotarea x from -5 to 3, y from -0.5 to 3
{\multiply\linethickness by 3 \putrule from -5 0 to 3 0 }
\putrule from 0 0 to 0 3
 \arrow <7pt> [.2,.67] from 0 0.5 to 1 0.5
 \put{$F$} [l] <2pt,0pt> at 1 0.5
\setdashes
\noindent\putrule from -5 1 to 0 1
\hshade 0  -5 0
        1  -5 0 /
\put{$x$} [br] <0pt,2pt> at 3 0
\put{$z$} [lt] <2pt,0pt> at 0 3
 \put{$E$} at -2.5 0.5
 \endpicture 
 }
\caption{\label{fig:testwall}The reflecting plate is in the 
 $x$--$y$ plane.
 The test wall is momentarily in the $z$--$y$ plane.
 The force on the test wall from the vacuum energy to its left is 
to be studied.
 See \sref{implications} for discussion of the ignoring of the 
right side of the test wall, and see \sref{fullbox} for discussion 
of the neglect of the finite size and other walls of the cavity on 
the left.} 
  \end{figure}

 To understand the nonvanishing pressures parallel to the wall,
imagine another planar boundary --- a ``test wall'' --- 
 at $x=0$ and  find the force on 
it (from the left side only --- see  \fref{fig:testwall}). 
 The volume of space occupied by boundary energy
 increases with $x$, so the total energy on the left 
 also changes linearly when the test wall moves.

 In accordance with the principle of energy balance 
 (also called principle of virtual work) one  expects 
\be
   F =\int_0^\infty p_1\,dz = -\, {\partial{U}\over\partial{x}}
  = - \int_0^\infty  \rho \,dz \equiv -E .
 \label{balance}\ee
 Here a trivial transverse dimension is suppressed in the 
notation: $F$ is a force per unit length, 
$E$ is an energy per unit area, and $U$ is the total energy per 
unit length,
equal to $E$ times the length of the cavity (which may be 
arbitrarily large).
If \emph{all the cutoffs are removed},  one has
$\displaystyle \rho = (32\pi^2 z^4)^{-1} = - p_1\,$,  
so energy balance is formally satisfied point-by-point in~$z$,
  but of course the integrals are divergent.

 Consider now the traditional ultraviolet cutoff, $t\ne 0$ but 
$\mathbf{r}_\bot=0$.  Evaluating the integrals in \eref{balance} 
with the integrands from \eref{wallrho} and~\eref{wallp}, one finds 
that
 \be
 F = +\,{\textstyle \frac12} E = {1\over 16\pi t^3}\,,
 \label{papabear}\ee
 a clear discrepancy with \eref{balance} in both sign and 
magnitude.
 (For details see the Appendix.)
 The $E$ defined in this way is the same as obtained from the 
second (surface) term in the small-$t$ expansion of 
\be
 E = -\,\frac12 \int_\Omega 
 \frac{\partial^2 \tbar}{\partial t^2}(t,\mathbf{r},\mathbf{r})\,d^3x =
\frac12 \sum_{n=1}^\infty \omega_n
e^{-t\omega_n}.
\label{totalE} \ee 
 But we shall argue that this $E$ is wrong, whereas the $F$ and 
\eref{balance} are correct (within the framework of the model).
 
 Next look at point-splitting perpendicular to the test wall
 ($x\ne0$, $t=0=y$).
 It is easy to see that the calculation is identical to the 
previous one except that $(t,\rho)$ change places with $(x,-p_1)$.
 Thus the result is 
 \be
 F = +2 E.
 \label{mamabear}\ee

 Finally, consider point-splitting in the ``neutral'' direction,
 $y\ne0$, $t=0=x$.
 One gets
 \be
F = -E,
\label{babybear}\ee
 as should happen, according to \eref{balance}.
In fact, the balance relation holds pointwise for the integrands;
  their values are
 \be
\rho=\frac1{2\pi^2 (y^2+4z^2)^2}\,, \qquad
 p_1= -\,\frac1{2\pi^2 (y^2+4z^2)^2}\,.
 \label{fullbabybear}\ee
The quantities in \eref{babybear} and \eref{fullbabybear} agree with 
\eref{papabear} for the pressure and with \eref{mamabear} for the 
energy.  We therefore propose to adopt them as 
``correct''\negthinspace.
 Note that for the Dirichlet wall, this $E$ is positive, unlike the 
traditional boundary energy from~\eref{totalE}.

A similar calculation for the ``conformal correction'' terms 
\eref{xirho}--\eref{xip} is entirely uneventful.
 They do not exhibit the anomaly:  $\Delta p_1 = -\Delta\rho$ always.
 Furthermore, as long as the cutoff is finite, their integrals over 
all~$z$ are $0$ anyway, as befits their origin in an integration by 
parts in a theory where the field vanishes on the boundary.
 (That this consistency relation fails without the cutoff was 
pointed out by Ford and Svaiter~\cite{FS}.)

 It may appear strange that the pressure and the energy have 
opposite sign.  This behavior is standard, however, for an energy 
density that remains constant as a geometrical parameter is varied, 
so that the total energy is proportional to the volume occupied.
 It holds, for instance, for the surface tension between two fluid 
media \cite[pp.~84--87]{pippard}, and also for cosmological ``dark 
energy'' (cf.~\eref{lambdaterm}).

\subsubsection*{Basic equations: general relativity.}
 In the notation of \cite{schutz}, the linearized Einstein equation 
is
 \be
\square \bar{h}_{\mu\nu}= -16\pi T_{\mu\nu} \,,
 \label{einst}\ee
 where
 \be
  \bar{h}_{\mu\nu} =  h_{\mu\nu}   -  
  \frac{1}{2}  ( \Tr \mathsf{h}  ) \eta_{\mu\nu}\,, 
 \qquad h_{\mu\nu} = g_{\mu\nu} - \eta_{\mu\nu} \,.
 \label{hdef}\ee
We apply \eref{einst} to our model of a quantized scalar field with 
a single plane boundary, the boundary condition being one of the 
point-splitting cutoffs discussed above, approaching the Dirichlet 
condition as a limit. 
Gravitational effects in the lab, although formally infinite in the 
theory without cutoff, are presumably actually tiny. 
 Therefore, the linearized Einstein equation (with a flat background) 
 should be physically 
adequate.  In a slightly better approximation one
  might use a curved background to represent the effect of the mass 
  of the boundary.

We assume a static solution.
 Then, tentatively adopting the usual 
ultraviolet cutoff, we have the basic equation of~\cite{Leipzig},
 \be
 -\nabla^2 \h_{00} = 16\pi\rho =
  \frac8{\pi}\,
\frac {4z^2-3t^2}{(t^2+4z^2)^3}\,\theta(z) + (\cdots)\theta(-z).
 \label{heq}\ee
 Because we consider only the region $z>0$ and the equation is linear,
  the term describing matter sources at negative~$z$ decouples and 
  can be ignored. However, it is essential to include the factor 
  $\theta(z)$ to study the singular limiting behavior of 
$\h_{00}$   and~$\rho$ at $z=0$.
For an infinite plane wall, $\h_{00}$ should depend only on~$z$, 
 so $\nabla^2 \h_{00} = \h_{00}''(z)$.

The solution of \eref{heq} that vanishes for $z<0$ is
\be
  \h_{00}(z) =  \frac{\theta (z)}{  \pi }  \biggl[  
\frac{4z}{t^3}
  \tan^{-1}   \biggl (\frac{2z}{t}  \biggr )
-\frac1{t^2+4z^2} + \frac1{t^2}\biggr ].
 \label{hsol}\ee
For a complete solution one should add a similar term with 
support in $z\le0$ representing the effect of the matter in that region,
  and also a possible global solution (affine linear in $z$) of the 
  homogeneous equation. (If the coefficient of $z$ in this homogeneous
   solution is made different on the two sides, a delta function
  representing the gravitational field of the wall itself will be added
  to~$\rho$.)

  Taking the limit $t\downarrow0$ in \eref{heq}  will yield a
differential equation 
with a distribution as its source.
  Taking that limit  in the solution will yield a singular distribution.
 Both limits involve somewhat arbitrary regularizations
  (Hadamard finite parts).
 Before presenting the details, we review some needed distribution 
theory in the next subsection.

\subsubsection*{Distributions and the moment expansion theorem.}

Our analysis requires the study of the behavior of several quantities of
physical interest as a parameter, $t,$ approaches $0.$ The singularity of the
quantity suggests that instead of an ordinary limit one needs to consider a
\emph{distributional }limit. Fortunately, a distributional theory of
asymptotic expansions is available \cite{EK}.

The key result in the distributional theory of asymptotic expansions is the
\emph{moment asymptotic expansion,} that says that if $f$ is a distribution of
one variable, $f\in\mathcal{D}^{\prime}\left(  \mathbb{R}\right)  ,$ that
\emph{decays rapidly} at $\pm\infty,$ then the asymptotic behavior of
$f\left(  \lambda x\right)  $ as the parameter $\lambda$ becomes large is%
\begin{equation}
f\left(  \lambda x\right)  \sim\sum_{n=0}^{\infty}\frac{\left(  -1\right)
^{n}\mu_{n}\delta^{\left(  n\right)  }\left(  x\right)  }{n!\lambda^{n+1}}\,.
\label{D.1}%
\end{equation}
Here the $\left\{  \mu_{n}\right\}  $ are the moments of the distribution $f,$
namely,%
\begin{equation}
\mu_{n}=\left\langle f\left(  x\right)  ,x^{n}\right\rangle \,, \label{D.2}%
\end{equation}
which become $\mu_{n}=\int_{-\infty}^{\infty}f\left(  x\right)  x^{n}dx$ if
$f$ is a locally integrable function or become $\mu_{n}=\sum_{k=-\infty
}^{\infty}a_{k}(b_{k})^{n}$ if $f\left(  x\right)  =\sum_{k=-\infty}^{\infty
}a_{k}\delta(x-b_{k})$ is a train of delta functions.

Observe that the behavior of $f$ at $\pm\infty$ is given by considering its
\emph{parametric} behavior; this reflects the fact that distributions do not
have point values, in general, as ordinary functions do. An alternative
approach that uses the ideas of Ces\`{a}ro summability is 
possible
\cite{est2},\ but in the present article it is exactly the parametric behavior
that we need to consider.

We used the term \textquotedblleft of rapid decay at $\pm\infty$%
\textquotedblright\ to describe the distributions for which the moment
asymptotic expansion holds. In technical terms, $f$ is of rapid decay at
$\pm\infty$ if it belongs to the distribution space $\mathcal{K}^{\prime
}\left(  \mathbb{R}\right)  ,$ dual of $\mathcal{K}\left(  \mathbb{R}\right)
.$ The space $\mathcal{K}$ is formed by the so-called GLS symbols \cite{7}; a
smooth function $\phi$ belongs to $\mathcal{K}$ if there is a constant
$\gamma$ such that $\phi^{(k)}(x)=O(|x|^{\gamma-k})$ as $|x|\rightarrow\infty$
for $k=0,1,2,\ldots,$ that is, if $\phi(x)=O(|x|^{\gamma})$ strongly; the
topology of $\mathcal{K}$ is given by the canonical seminorms.

Notice that if $\mathcal{E}^{\prime}\left(  \mathbb{R}\right)  $ is the space
of distributions with compact support, then $\mathcal{E}^{\prime}\left(
\mathbb{R}\right)  \subset\mathcal{K}^{\prime}\left(  \mathbb{R}\right)  ,$ so
that the moment asymptotic expansion holds for distributions of compact support.

If $f$ is a distribution of one variable that does not belong to
$\mathcal{K}^{\prime}\left(  \mathbb{R}\right)  $ then it does not satisfy the
moment asymptotic expansion, and, in general, the behavior of $f\left(
\lambda x\right)  $ as $\lambda\rightarrow\infty$ is very complicated. There
is one situation, however, when $f\left(  \lambda x\right)  $ has a
simple expansion~\cite{r43}.
 Indeed, suppose that $f$ has support bounded on the left,
$\mathop{\rm supp}f\subset\lbrack a,\infty)$ for some $a$, 
and suppose that
$f$ is an ordinary locally integrable function for $x$ large and that the
\emph{ordinary} expansion%
\begin{equation}
f\left(  x\right)  =b_{1}x^{\beta_{1}}+\cdots+b_{n}x^{\beta_{n}}+O\left(
x^{\beta}\right)  \,,\quad \hbox{as 
}x\rightarrow\infty\,,\label{D.3}%
\end{equation}
holds, where $\beta_{1}>\beta_{2}>\cdots>\beta_{n}>\beta$ and where $-\left(
k+1\right)  >\beta>-\left(  k+2\right)  $ for some integer $k.$ Then $f\left(
\lambda x\right)  $ has the distributional development%
\begin{equation}
f\left(  \lambda x\right)  =\sum_{j=1}^{n}b_{j}g_{j}\left(  \lambda x\right)
+\sum_{j=0}^{k}\frac{\left(  -1\right)  ^{j}\mu_{j}\delta^{\left(  j\right)
}\left(  x\right)  }{j!\lambda^{j+1}}+O\left(  \lambda^{\beta}\right)
\,,\label{D.4}%
\end{equation}
as $\lambda\rightarrow\infty.$ Here the distributions $g_{j}\left(  x\right)
$\ are suitable regularizations of the functions $\theta\left(  x\right)
x^{\beta_{j}},$ namely $\mathcal{P}f\left(  \theta\left(  x\right)
x^{\beta_{j}}\right)  ,$\ as we explain below, while the $\mu_{j}$ are still
the moments, but understood in a \emph{generalized} sense, since the integrals
$\int_{a}^{\infty}f\left(  x\right)  x^{j}dx$ will be divergent at infinity,
in general, and thus we need to consider their Hadamard finite part,
$\mathrm{F.P.}\int_{a}^{\infty}f\left(  x\right)  x^{j}dx.$

Interestingly, the asymptotic expansion (\ref{D.4}) considers the behavior of
$f$ at infinity, but it is writen in terms of distributions that could be
singular at the origin. The process of associating a distribution to a
singular, not locally integrable, function is considered in 
\cite{EF2001}; in
the mathematical literature one says that the distribution is a
\emph{regularization} of the singular function. There are many regularization
procedures, and, in general, regularizations are not unique \cite{E03}; for
our purposes we need to consider the Hadamard finite part of integrals and the
corresponding regularization method.

Let us first recall that the \emph{finite part of the limit} of $F\left(
\varepsilon\right)  $ as $\varepsilon\rightarrow0^{+}$ exists and equals $A$
if we can write $F\left(  \varepsilon\right)  =F_{\mathrm{fin}}\left(
\varepsilon\right)  +F_{\mathrm{infin}}\left(  \varepsilon\right)  ,$ where
the \emph{infinite} part $F_{\mathrm{infin}}\left(  \varepsilon\right)  $ is a
linear combination of functions of the type $\varepsilon^{-p}\ln
^{q}\varepsilon,$ where $p>0$ or $p=0$ and $q>0,$ and where the \emph{finite}
part $F_{\mathrm{fin}}\left(  \varepsilon\right)  $ is a function whose limit
as $\varepsilon\rightarrow0^{+}$ is~$A$ \cite{EK}. We then 
write%
\begin{equation}
\mathrm{F.P.}\lim_{\varepsilon\rightarrow0^{+}}F\left(  \varepsilon\right)
=A\,.\label{D.5}%
\end{equation}
A similar idea can be applied to limits at infinity, which turns out to be
equivalent to a change of variables,%
\begin{equation}
\mathrm{F.P.}\lim_{\lambda\rightarrow+\infty}G\left(  
\lambda\right)
=\mathrm{F.P.}\lim_{\varepsilon\rightarrow0^{+}}G\left(  1/\varepsilon\right)
\,.\label{D.6}%
\end{equation}

Suppose now that we have an integral $\int f\left(  x\right)  dx$ such that
$\int_{c}^{b}f\left(  x\right)  dx$ is well defined for $c>a,$ but for which
$\int_{a}^{b}f\left(  x\right)  dx$ is divergent. Then one considers the
function $F\left(  \varepsilon\right)  =\int_{a+\varepsilon}^{b}f\left(
x\right)  dx$ for $\varepsilon>0,$ and if the finite part of the limit exists
as $\varepsilon\rightarrow0^{+},$ then we call this value the \emph{Hadamard
finite part }of the integral, and write%
\begin{equation}
\mathrm{F.P.}\int_{a}^{b}f\left(  x\right)  dx=\mathrm{F.P.}\lim
_{\varepsilon\rightarrow0^{+}}\int_{a+\varepsilon}^{b}f\left(  x\right)
dx\,.\label{D.7}%
\end{equation}

Let $f$ now be a function of one variable that vanishes for $x<a,$ that is
locally integrable in the open interval $\left(  a,\infty\right)  ,$ but for
which the integral $\int_{a}^{b}f\left(  x\right)  dx$ is divergent for $b>a.$
Then, in general, one cannot associate a distribution in 
$\mathcal{D}^{\prime
}\left(  \mathbb{R}\right)  $ to $f.$ Nevertheless, if the Hadamard finite
part of the integral, $\mathrm{F.P.}\int_{a}^{b}f\left(  x\right)  
dx$, exists
for $b>a$, then one can define a distribution $\mathcal{P}f\left(  
f\right)  ,$
its \textquotedblleft partie 
finie\textquotedblright\negthinspace, by 
putting%
\begin{equation}
\left\langle \mathcal{P}f\left(  f\left(  x\right)  \right)  ,\phi\left(
x\right)  \right\rangle =\mathrm{F.P.}\int_{a}^{b}f\left(  x\right)
\phi\left(  x\right)  dx\,,\ \ \ \ \phi\in\mathcal{D}\left(  \mathbb{R}%
\right)  \,.\label{D.8}%
\end{equation}

In the case when $f\left(  x\right)  =\theta\left(  x\right)  x^{\beta}$ for
some $\beta\in\mathbb{C}$ we obtain the distribution $\mathcal{P}f\left(
\theta\left(  x\right)  x^{\beta}\right)  .$ When $\Re e\,\beta>-1,$ however,
the function $\theta\left(  x\right)  x^{\beta}$ is locally integrable even at
$x=0$ and thus $\mathcal{P}f\left(  \theta\left(  x\right)  x^{\beta}\right)
$ reduces to the standard distribution associated to $\theta\left(  x\right)
x^{\beta},$ which is usually denoted by $x_{+}^{\beta};$ one can perform
analytic continuation of $x_{+}^{\beta},$ and obtain a distribution for
$\beta\neq-1,-2,-3,\ldots,$ and it can be shown \cite{EK}\ that
$\mathcal{P}f\left(  \theta\left(  x\right)  x^{\beta}\right)  =x_{+}^{\beta}$
for such values of $\beta.$ On the other hand, the symbol $x_{+}^{\beta}$ does
not make sense for $\beta=-1,-2,-3,\ldots,$ while $\mathcal{P}f\left(
\theta\left(  x\right)  x^{-k}\right)  $ is well defined for $k=1,2,3,\ldots.$

The formulas%
\begin{equation}
\left(  x_{+}^{\beta}\right)  ^{\prime}=\beta x_{+}^{\beta-1},\label{D.9}%
\end{equation}%
\begin{equation}
\left(  \lambda x\right)  _{+}^{\beta}=\lambda^{\beta}x_{+}^{\beta
}\,,\label{D.10}%
\end{equation}
hold for $\beta\neq-1,-2,-3,\ldots$ since they hold for $\Re e\,\beta$ large,
by the principle of analytic continuation. However, we have the modified
formulas%
\begin{equation}
\frac{d}{dx}\left(  \theta\left(  x\right)  \ln x\right)  =\mathcal{P}f\left(
\frac{\theta\left(  x\right)  }{x}\right)  \,,\label{D.11}%
\end{equation}%
\begin{equation}
\frac{d}{dx}\left(  \mathcal{P}f\left(  \frac{\theta\left(  x\right)  }{x^{k}%
}\right)  \right)  =-k\mathcal{P}f\left(  \frac{\theta\left(  x\right)
}{x^{k+1}}\right)  +\frac{\left(  -1\right)  ^{k}\delta^{\left(  k\right)
}\left(  x\right)  }{k!}\,,\label{D.12}%
\end{equation}
and%
\begin{equation}
\mathcal{P}f\left(  \frac{\theta\left(  \lambda x\right)  }{\left(  \lambda
x\right)  ^{k}}\right)  =\frac{1}{\lambda^{k}}\mathcal{P}f\left(  \frac
{\theta\left(  x\right)  }{x^{k}}\right)  +\frac{\left(  -1\right)  ^{k}%
\ln\lambda\,\delta^{\left(  k\right)  }\left(  x\right)  }{k!\lambda^{k}%
}\,.\label{D.13}%
\end{equation}

\subsubsection*{Distributional limits of energy and pressure.}
 In \cite{Leipzig} the foregoing mathematical theory was applied to 
\eref{heq} and \eref{hsol}.
 Let $\lambda=1/t$.  The limit of \eref{heq} is
\begin{eqnarray}
- \,\frac{ d^2 \h_{00}(z)}{dx^2} &=
 \frac{1 }{2\pi} \, \mathrm{ F.P.}
  \left(\frac   {\theta (z)}{z^4} \right)
- 2 \lambda^3 \delta (z) + \frac{ 1 }{\pi} \lambda^2 \delta ' (z)
\nonumber \\
&{} + \frac{ 1 }{8\pi} \delta ''' (z) 
 - \frac{ 1 }{12 \pi} \ln (2 \lambda ) \delta ''' (z) 
 + \mathcal{O}\biggl ( \frac{ 1 }{\lambda} \biggr ) ,
\label{heqlim}\end{eqnarray}
 and the limit of \eref{hsol} is
\begin{eqnarray}
 \h_{00}(z) &= 2
\lambda^3 \theta (z) z - \frac{ 1 }{\pi }\lambda ^2 \theta (z) -
\frac{ 1}{12 \pi} \, \mathrm{F.P.} \left( \frac{\theta (z)} 
{z^2} \right)
\nonumber \\
&{} - \frac{ 1 }{18\pi} \delta ' (z) + \frac{ 1 }{12 \pi} \ln
(2\lambda) \delta ' (z)
  + \mathcal{O}\biggl( \frac{ 1 }{\lambda} \biggr).
\label{hsollim}\end{eqnarray}
(The apparent dimensional incoherence of the $\delta'\, 
\ln(2\lambda)$ term
 is attributable to the violation of scale invariance by the 
finite-part operation.)

One quickly verifies that \eref{hsollim} is a solution of 
\eref{heqlim}.
 \emph{The limit of the solution is the solution of the limit.}
 This also follows from the definitions of the distributional 
operations.

 A consequence of the moment expansion theorem is that,
 when one uses a singular function to define a distribution,
  divergent leading powers can be 
replaced by derivatives of $\delta$ with 
\emph{arbitrary finite} coefficients~\cite{EF2001}.
These terms can be interpreted as renormalizations of properties of 
the boundary.
The previous observation thus shows that our toy Einstein equation survives 
 the renormalization process
as a mathematically consistent differential equation.

We shall now go beyond \cite{Leipzig} in two ways.
 First, keeping $t$ as the cutoff parameter, we examine the 
differential equation for the pressure component, using
 \eref{wallp} with $x=0=y$:
\begin{equation}\label{hpeq}
- \,\frac{ d^2   \h_{11}(z)  }{d z^2 } = 16 \pi p_{1} =
  -\, \frac{ 8 }{\pi }\,  \frac1{(t^2 + 4 z^2 )^2} \theta(z).
\ee 
The appropriate   solution analogous to \eref{hsol} is 
\begin{equation}
\h_{11}(z) =  \frac{  \theta (z)  }{\pi} \biggl[ \frac{2z}{t^{3}} 
  \tan^{-1}  \biggl ( \frac{2 z }{ t } \biggr )  \biggr  ] .
\label{hpsol}\end{equation}

 For the evaluation of the asymptotic behavior of \eref{hpeq}, 
 the relevant distribution to use in the moment expansion theorem is 
\begin{equation}
f_{1}  (z )  = \frac{1 }{ (1  +  4 z^2 )^2 } \theta(z)  = 
\frac{  1}{ 16}  \frac{1 }{ z^4 }  + \mathcal{O} \biggl (
  \frac{ 1 }{ z^6 }\biggr )   \mathrm{~as~} 
 z\to \infty.
\end{equation}
The moment expansion theorem states, up to the relevant order, 
 that the asymptotic expansion of $f_{1}  (\lambda z)$ is 
\begin{eqnarray}
f_{1}   (\lambda z )  &  \sim \frac{ 1 }{16 }   \mathcal{P}f
    \biggl ( \frac
{ \theta (\lambda z )   }{   (\lambda z)^4  }  \biggr )  
   + \sum_{j= 0}^{3}
 (- 1 )^j \mu_{j }   (f_{1}  ) \frac{ \delta^{( j )} 
 (\lambda z)}{     j ! }
  + \mathcal{O}  \biggl (\frac{ 1 }{ \lambda^{5} }  \biggr ) 
\nonumber \\
& = \frac{ 1 }{16 } \biggl \lbrace \frac{ 1 }{ \lambda^4 } 
    \mathcal{P}f \biggl 
( \frac{ \theta ( z )   }{    z^{4}  }  \biggr ) - \frac{ \ln \lambda}
{ 3! \lambda^{4}  } \delta^{'''}(z) \biggr \rbrace  \nonumber \\
&{}+  \sum_{j = 0}^{ 3 }  (- 1 )^{j}  \mu_{j }  ( f_{1}  ) 
\frac{  \delta^{  ( j  ) } (z  ) }{ j!  \lambda^{j+1 } }  + \mathcal O 
\biggl ( \frac{ 1 }{ \lambda^{5} } \biggr ), 
\end{eqnarray}
where the moments $\mu_{j}( f_{1})$ of the function $f_{1}$ are
\begin{equation}
\mu_{0}(f_{1} ) = \int_{0}^{\infty }  \frac{ 1 }{ (  1 + 4 z^2 )^2 } \, dz
  = \frac{ \pi }{ 8} \,,
\end{equation}
\begin{equation}
\mu_{1}  (f_{1}  ) = \int_{0}^{\infty }  \frac{ 1 }{ (  1 + 4 z^2 )^2 } 
 \, z \, dz = \frac{ 1  }{ 8} \,,
\end{equation}
\begin{equation}
\mu_{2}  (f_{1}  ) = \int_{0}^{\infty }  \frac{ 1 }{ (  1 + 4 z^2 )^2 }
  \, z^{2} \, dz = \frac{ \pi  }{ 32}\, ,
\end{equation}
\begin{equation}
\mu_{3} ( f_{1} ) = \mathrm{F.P.} \int_{0}^{\infty } 
  \frac{ 1 }{ (  1 + 4 z^2 )^2 } \, z^{3} \, dz 
 =- \frac{ 1}{32} + \frac{ 1 }{16} \ln 2 . 
\end{equation}
Therefore,                  
the distributional limit of the differential equation \eref{hpeq} is 
\begin{eqnarray}
- \,\frac{ d^2  \h_{11}(z)  }{d z^{2}  } & = -\, \frac{  1 }{ 2 
\pi } 
\mathcal{P}f \biggl ( \frac{ \theta ( z )   }{   z^4  }  \biggr )
   - \lambda^{3} 
 \delta(z) + \frac{  1  }{  \pi } \lambda^{2} \delta^{' } (z) 
 - \frac{ 1 }{ 8 }  \lambda  \delta^{''} (z)   \nonumber \\
&{}   - \frac{ 1 }{ 24 \pi }  \delta '''  (z )  
 +  \frac{ 1 }{ 12 \pi }\ln (2 \lambda ) \delta^{'''}(z) 
  +\mathcal{O} \biggl ( \frac{ 1 }{ \lambda } \biggr ).  
  \label{hpeqlim}\end{eqnarray}

 Similarly,
the relevant function for the analysis of \eref{hpsol} is
\begin{equation}
f_{2} ( z )    = z \tan^{-1}   (2 z )  \theta(x)   
 = \frac{   \pi }{ 2 }   z - \frac{ 1}{2 }  
 + \frac{ 1 }{ 24 } \frac{ 1}{ z^2}  +\mathcal{O}
  \biggl ( \frac{ 1 }{ z^{4} }  \biggr) .
\end{equation}
The moment expansion theorem says that
\begin{eqnarray}
 f_{2} (\lambda z )  &\sim   \frac{   \pi }{ 2 }  \theta ( \lambda z) 
  (\lambda z ) - \frac{ 1}{2 } \theta (\lambda z)  + \frac{ 1 }{24 }
 \mathcal{P}f   \biggl ( \frac{ \theta (\lambda z )   }
 {   (\lambda z)^2  }  \biggr )  \nonumber\\
 &{} +  \sum_{j = 0}^{  1 }  (-1)^{j} \mu_{ j } (f_{2} )
   \frac{ \delta^{(j)  } ( \lambda z)  }{ j! } + \mathcal{O}
      \biggl (  \frac{1 }{\lambda}\biggr)  \nonumber\\
 & = \frac{ \pi }{ 2 } \lambda  \theta (z) z - \frac{  1}{ 2 }
  \theta(z )   + 
 \frac{  1 }{ 24 \lambda^2} \mathcal{P}f
    \biggl ( \frac{ \theta ( z )   }{   z^2  }  \biggr ) 
 - \frac{ 1}{ 24}  \frac{ 1 }{ \lambda^2} \ln \lambda \delta^{'}( z ) 
  \nonumber \\
 &{} + \sum_{j = 0}^{  1 }  (-1)^{j} \mu_{ j } (f_{2} ) 
  \frac{ \delta^{(j)  } (  z)  }{ j! \lambda^{j +1 } }
  + \mathcal{O}     \biggl (  \frac{1 }{\lambda}\biggr)  . 
\end{eqnarray}
Finally, the relevant moment expansion coefficients this time are
\begin{equation}
\mu_{0}   ( f_{2}  )    = \int_{0}^{\infty}  z \tan^{-1} 
   (2 z )\, d z = \frac{ \pi }{ 16 }\,,
  \end{equation}
\begin{equation}
\mu_{1 }( f_{2} )  = \int_{0}^{\infty}  z^2 \tan^{-1}   (2 z )\, 
d z
  = \frac{ 1}{ 72 } + \frac{ 1 }{ 24 } \ln 2.
\end{equation}
 Forming the correct linear combination of these terms, we have
\begin{eqnarray}
\h_{11}(z) & =  \lambda^3 \theta(z)z  -\frac{ 1}{ \pi }  
\lambda^2
  \theta(z) 
 + \frac{ 1}{ 12 \pi } \mathcal{P}f   \biggl ( \frac{ \theta ( z ) }
 {   z^2  }  \biggr ) + \frac{ 1 }{8} \lambda \delta (z)  
\nonumber\\
&{} - \frac{2 }{ \pi }  \biggl [ \frac{ 1}{ 72 } + \frac{ 1 }{ 24 }
  \ln 2 \biggr ] \delta^{'}(z)  - \frac{1}{ 1 2 \pi } \ln \lambda 
 \delta^{'}(z) 
 +\mathcal{O} \biggl ( \frac{ 1 }{ \lambda } \biggr ) \nonumber\\
& =  \lambda^3 \theta(z)z  - \frac{ 1}{ \pi }  \lambda^2 \theta(z) 
 + \frac{ 1}{ 12 \pi } \mathcal{P}f   \biggl ( \frac{ \theta ( z )}
 {z^2  }  \biggr ) + \frac{ 1 }{8} \lambda \delta (z) 
 - \frac{ 1}{ 36 \pi } \delta^{'}(z) \nonumber \\ 
 &{}- \frac{ 1 }{ 12 \pi} \ln ( 2 \lambda ) \delta^{'}(z)
   +\mathcal{O} \biggl ( \frac{ 1 }{ \lambda } \biggr ).   
\label{hpsollim}\end{eqnarray}
   
By taking the second derivative of the solution \eref{hpsollim},
  according to the Hadamard $\mathcal {P}f$ formulas, 
we verify that the differential equation \eref{hpeqlim} 
is satisfied.
 This completes the counterpart for the pressure of the energy 
calculation in~\cite{Leipzig}. 

 Next we want to see what happens when we use $y$ as the cutoff 
parameter.  It is neutral with respect to both the time--energy 
and the $x$--$p_1$ dimensions.  The pertinent formulas 
are~\eref{fullbabybear}. One immediately sees that the $p_1$ 
calculation is identical to what we just did, with $y$ in the role 
of~$t$, and also that $\rho$ in this case is just the negative 
of~$p_1\,$.
 So we do not need to do any more calculating, and we have three 
new, agreeing results (two for $p_1$ and one for $\rho$)
 that outvote the old 
formula for~$\rho$   in \cite{Leipzig} and \eref{heq}.
 Similarly, we have formulas for $\h_{11}$ and $\h_{00}$
that now seem more trustworthy than the old formula for 
$\h_{00}$ in \cite{Leipzig} and \eref{hsol}.

 \subsubsection*{Comparison of old and new energy formulas.}
 The old formula \eref{hsollim} for  the distributional limit
  of $\h_{00}$, based on the $t$~cutoff, was
\begin{eqnarray}
\h_{00}^{\mathrm{old}} (z)  & =  2 \lambda ^3 \theta (z)  z  
-\frac1{ \pi } \lambda^2  \theta (z) - \frac1{ 12 \pi } 
 \mathcal{P}f  \biggl ( 
\frac{  \theta (z )  }{ z^2 }  \biggr ) \nonumber \\
& {}-  \frac{  1 }{ 18 \pi }  \delta ' (z)  
 + \frac1{ 12 \pi } \ln ( 2 \lambda ) \delta ' (z ).
\label{solutionhold} \end{eqnarray}
The new one, based on the $y$~cutoff,
  is the negative of \eref{hpsollim}:
\begin{eqnarray}
\h_{00}^{\mathrm{new}}(z)  & = - \lambda^3 \theta(z)z  
 +\frac{1}{ \pi }  \lambda^2 \theta(z) 
 - \frac{ 1}{ 12 \pi } \mathcal{P}f \biggl ( \frac{ \theta ( 
z )   }{   z^2  }  \biggr ) - \frac{ 1 }{8} \lambda \delta (z) 
  \nonumber\\ 
 & {}+ \frac{ 1}{ 36 \pi } \delta^{'}(z) 
 + \frac{ 1 }{ 12 \pi} \ln ( 2 \lambda ) \delta^{'}(z) . 
\label{solutionhnew} \end{eqnarray}
The structure is exactly the same (except for the accidental 
absence of a $\delta(z)$ term from \eref{solutionhold}).
 Only the numerical coefficients are different;
 their ratios are simple integers, shown in \tref{ratio}.
 The coefficients are the same for the regularized 
 $\theta(z)/z^2$ term, which is the only term describing a spatially
 extended vacuum effect of the quantum field theory.
 All the other terms  (apart from the bare $\delta'(z)$ terms, 
which could in each case be absorbed by rescaling the argument of 
the logarithm) involve powers or logarithm of the large cutoff 
parameter~$\lambda$.
 The spirit of renormalization is to regard these terms as part of 
the wall, not the field, and to replace the $\lambda$-dependent 
factors by arbitrary finite constants. 
 Note that the two terms involving $\theta(z)$ can be regarded as 
part of the solution of the homogeneous Einstein equation, and the 
other two terms are localized on the wall.
A full discussion of the homogeneous solution is beyond the scope
 of this paper.

 \Table{\label{ratio}Ratio 
$h_{00}^{\mathrm{old}}/h_{00}^{\mathrm{new}}$.}
\br 
\hfil{}Term\hfil&Ratio  of     coefficients\\ 
\mr
 \hfil$\lambda ^3  z  \theta (z) $\hfil&\hfil$-2$\hfil\\
 \hfil$\lambda^2 \theta (z) $\hfil&\hfil$-1$\hfil\\ 
 \hfil$\lambda \delta(z) $\hfil &\hfil0\hfil  \\
 \hfil$ \ln (2 \lambda )  \delta ' (z )$\hfil&\hfil1\hfil \\ 
\hfil$ \delta ' (z)  $\hfil&\hfil$-2$\hfil\\ 
  \hfil$\mathcal{P}f \left(\displaystyle \frac{ \theta (z ) }{ z^2  }
   \right)$\hfil&\hfil1\hfil  \\ 
 \br 
\endtab

\section{Caveats and implications:  A dialogue}  
\label{implications}
{ \advance\parskip 1\jot

 \salv Friends, I am perplexed by the paradoxical behavior of the 
energy and pressure in my model.
 The change in energy as the side of a box moves does not match the 
force on the side.  Even the sign is wrong!

 \simp I don't understand why you are so upset.  We know that the 
divergent terms in vacuum energy are unphysical and must be 
discarded. The divergences you're looking at are not logarithmic, 
so in an analytic regularization scheme (dimensional 
regularization or zeta functions) they do not arise in the first 
place.  Also, those methods are Lorentz-covariant, whereas 
 point-splitting in any particular direction is not; it's not 
surprising that the latter gives noncovariant results for the 
cutoff-dependent terms.

 \sagr I disagree with Simplicio about the correct way to think 
about divergences in Casimir calculations. 
 (With a realistic boundary the ``divergences'' will be finite but 
nonzero, and they have physical content.)
  But I, too, think that 
your ``paradox'' is a nonproblem, for a whole slew of reasons.
 First, in the calculation symbolized by \fref{fig:testwall}
  you have 
completely ignored the space to the right of the test wall.  
 Surely it is obvious that the force $F$ is cancelled by an equal 
and opposite force from the other side.
 Second, your scalar field is a toy that doesn't exist in the real 
world.  Nature's massless field is electromagnetism, and one of the 
most famous properties of electromagnetic Casimir energy is that 
the surface divergences you're studying vanish there,  because 
the electric and magnetic contributions cancel. 
Also, there is no surface divergence for a flat boundary and a 
scalar field if $\xi=\frac16$.
  Third, 
 your cutoff is not a physical model of a realistic boundary;
 it is just an ad hoc procedure.
 If you use a noncovariant or otherwise unphysical cutoff,
  you must be prepared to 
insert unphysical counterterms; they are not a physical pathology 
but an artifact of a suboptimal formulation of the problem.

 \salv Give me a few minutes to respond to this barrage!
First, this anomaly was first noticed in calculations for a thin 
reflecting spherical shell
 [S. Fulling and M. Schaden, unpublished].
 In that case the force in question acts on the sphere itself,
  not a test surface perpendicular to it. 
   The leading term in the surface energy per unit area is 
the same for convex and concave surfaces, and the area (hence the 
volume occupied by the surface energy) depends on radius in the 
same way. 
So from the energy point of view, the inside and outside forces 
 should be in the same direction!
  And indeed, the direct calculation of force from pressure gave 
 that result,
but with that same discrepant factor, $-\frac12$ \cite{safqfe09}.

 \simp Why not just show us the sphere calculations?

 \salv  They are rather complicated (and computer-assisted) and not 
yet available in publishable form.  Instead, I have shown you that 
the same phenomenon occurs in the much simpler rectangular 
configuration.   My belief is that energy-pressure balance ought to 
hold for each side of the test wall separately, since they are 
effectively independent systems (especially if the test wall is 
itself another perfectly reflecting wall).  I don't know or care 
what exists on the other side of the wall.  But if you insist that 
I analyze a thin partition, taking both sides into account, I can 
imagine scenarios where the two sides don't balance.  Recall that 
the surface energy for a Neumann plate has the opposite sign from a 
Dirichlet plate.  So if the plate in \fref{fig:testwall} has the 
 Neumann property for $x>0$, 
 the resulting force on the test wall will be in 
the same direction as that from the Dirichlet plate on the other 
side.

 \sagr Whoa!  Does your plate magically change from one material to 
another as the test wall moves?

 \salv  At the instant portrayed in \fref{fig:testwall}, the force is 
as I've described.  What happens after the test wall has moved a 
finite distance is not really relevant. 
 But anyway, I said I can ``imagine'' these situations, not create 
them in a lab.  Perhaps the reflecting media stretch and shrink as 
the test wall is moved, but in such a way that their reflective
 properties don't change.  I see no reason why such a situation 
could not exist in principle; we're talking here about preserving a 
very general physical principle.
 Or perhaps the Neumann material is a thin strip that winds around 
a spool  as the test wall moves to positive~$x$.

 \sagr  But the boundary of the Neumann substance still exists even 
if it is wound into a spiral.  This reminds me of your somewhat 
ill-fated pistol design~\cite{rect}.
 Leaving aside the fact that the pistol sucked in its bullet 
instead of firing it, you had to admit that two conductors in close 
contact are not the same thing as a solid block of conducting 
material; rather, one has a limiting case of a small gap between 
conductors, and in the gap the Casimir energy is very important.

 \salv Yes, but our point was that the idealized theory simply must 
break down when the gap is very small (atomic dimensions),
  even if we don't know exactly how.  I admit that the spool 
  scenario has complicating factors.  How about this one:
 There is no boundary surface at all along the $z$~axis at 
positive~$x$.  Instead, (to use electrical terminology) the right 
side of the apparatus is filled with a liquid conductor --- 
mercury --- which is pushed out of the way in the $z$ direction 
into a large tank as the test wall moves.  Then there is no 
boundary vacuum energy in the right half.
 My point is that there is a vast variety of things that could 
exist on the right of the wall, and no way to prove that they all 
give the same force on the wall.  Anyway, this rectangular scenario 
is just a simple model of a more general problem that can't be 
argued away.  I have already reported our result about the sphere.
 My reply to the point about electromagnetic surface energy is 
similar:  The  leading surface terms cancel there, but for curved 
surfaces there are higher-order terms that don't, and the same 
issue will probably arise.

 \simp You said something that bothered me.  You want to make your 
test wall into a perfect reflector in its own right.  So doesn't 
its vacuum energy need to be counted?  Also, you started by calling 
your apparatus a ``box''\negthinspace, implying that it has walls 
on all sides instead of extending to infinity as it does in your 
figure.  But then your argument for the correctness of the equation 
$p_3=0$ breaks down.  You argued that moving the boundary at $z=0$ 
would merely push the vacuum energy around instead of changing its 
total value.  That isn't true if the cavity's length is changing.
 And your argument for \eref{balance} explicitly assumed that the 
cavity's length in the $x$ direction \emph{is} finite and changing!
 You can't have it both ways.

 \salv  You are right. For a convincing and complete treatment we 
should consider a bounded system --- a rectangular box --- 
 and study the energy and pressure associated with all those walls.
 In fact, that has been done (see \sref{fullbox}).
 The analogue of the $p_3$ argument does work out, but 
contributions from infinitely many image sources must be combined 
to get energy-pressure balance.

 \sagr How seriously do you take your model physically?

 \salv I don't know.  There are two philosophies one could take.
First, one might hope that the finite-cutoff theory 
(possibly including some finite, physically motivated 
counterterms)
could be a 
physically plausible model of a real boundary, including its 
gravitational effects, at least at distances not extremely close to 
the wall.  The prototype I have in mind is the ``hard core'' in the 
Lennard--Jones potential, which nuclear physicists regard as 
adequate when they don't need to probe the chromodynamical 
structure of nucleons at short distances.
 Surely in that situation a theory that violates energy-pressure 
balance is unacceptable.
 Neutral point-splitting gives a theory that is acceptable in this 
respect, but I can't regard it as a logically sound, long-term 
solution, since it amounts to changing the rules of the game in 
each situation until I get a result that I like.  Its only 
justification is that, unlike less contrived alternatives, it does 
not immediately produce results that are obviously wrong. 
In the long run, detailed study of the physics of real boundaries 
will be necessary.

 The other point of view  is that the cutoff is a mathematical
  means to an end, which is a limiting theory with the cutoff 
 removed.  At the intermediate 
    stage a violation of energy-pressure balance may be 
    tolerated, so long as the final theory is physically 
    acceptable.  As you both remarked, an unphysical
  regularization can necessitate unphysical 
    counterterms, so that the undesirable terms appear in the final 
    equations with coefficient~$0$ (i.e., not at all). 
I could live with that, but it is certainly highly unaesthetic.

 \simp So, you have left us all with several questions to think about.
 }
  \begin{itemize}
\item Are both of those philosophies physically tenable?
 Or one, or neither?
 \item What is the meaning of terms proportional to $\delta$ and 
$\delta'$ in a component of a metric tensor?
 Do we learn anything more than if we had just discarded them?
 \item When you put in numbers, is the vacuum stress sufficiently small
 to justify your use of linearized Einstein equations?
 Must you add counterterms to assure that?
 \item In particular, I am concerned by the terms in $\h_{\mu\mu}$ that
 are linear in~$z$.
  What is the significance of the linear and constant terms (solutions
  of the homogeneous equation)?  To what extent can one exclude them by
  boundary conditions or remove them by gauge (coordinate)
   transformations?
 \item  Naively I would expect  a term that is independent 
 of~$\lambda$ to be ``physical'' and hence well defined.
But there is one such term in \tref{ratio} for which the ratio is not 
unity.
 Should we be worried about that? I can see that that term could be 
regarded as part of the logarithmic term, although the coefficient 
ratio for the logarithmic term itself \emph{is} unity.
\item What will you do near the corner of a parallelepiped,
 where there is no neutral direction?
 \end{itemize}

\section{Energy density and pressure inside a rectangular box}
\label{fullbox} 
 
  In this section we give a more careful accounting of the total 
  energy associated with two perpendicular boundaries as in 
  \fref{fig:testwall}.
 The test wall itself is made into a perfectly reflecting 
boundary, and  to make all integrals proper two other such walls 
are introduced parallel to the original ones.
 The cavity is thus a finite box as shown in \fref{fig:images}.
 More precisely, the cavity is an infinite rectangular waveguide, 
because we have no need to introduce boundaries parallel to the 
$x$--$z$ plane.  Therefore, in the following, ``energy'' should be 
read as ``energy per unit length'' (in the $y$ direction), and 
similarly for ``force''\negthinspace.

 The wave equation in this space can be solved by a sum over 
infinitely many images.  
 The analysis follows rather closely the two-dimensional theory in 
\cite{rect,Leipzig}, but the details of the expressions are 
different.
 As shown in \fref{fig:images}, there are four classes of images.
 Points periodically displaced 
 (with periods twice the length, $a$, and width, $b$, of the box) 
 from the source point at $\mathbf{r}'= (x',0,z')$ contribute to 
the cylinder kernel $\tbar(t,\mathbf{r},\mathbf{r}')$ the terms
 \be
\tbar_{\mathrm{P}mn} = -\,\frac1{2\pi}\,
 \frac1{t^2+y^2 +(2ma +x-x')^2 +(2nb+z-z')^2}\,.
\label{Pterms} \ee
 The term $\tbar_{\mathrm{P}00}$ is the free kernel $\tbar_0$ and 
hence should be omitted as it was in~\eref{Twall}.
 Points reflected horizontally (i.e., reflected through a vertical 
line in the lattice of \fref{fig:images}), together possibly with 
 a periodic displacement in the vertical direction, contribute
 \be
\tbar_{\mathrm{H}mn} = +\,\frac1{2\pi}\,
 \frac1{t^2+y^2 +(2ma -x-x')^2 +(2nb+z-z')^2}\,.
\label{Hterms} \ee
 Similarly, vertical reflections produce
 \be
\tbar_{\mathrm{V}mn} = +\,\frac1{2\pi}\,
 \frac1{t^2+y^2 +(2ma +x-x')^2 +(2nb-z-z')^2}\,.
\label{Vterms} \ee
Finally, reflections in both dimensions give rise to terms
 \be
\tbar_{\mathrm{C}mn} = -\,\frac1{2\pi}\,
 \frac1{t^2+y^2 +(2ma -x-x')^2 +(2nb-z-z')^2}\,.
\label{Cterms} \ee
(Such a point is connected to the source point by a straight line
that passes through an intersection of the horizontal and vertical 
lines of the lattice.  When folded back into the box, this line can 
be characterized as a classical path or optical ray that is 
reflected from a corner of the box. All the other terms have 
similar classical-path interpretations.)

 \begin{figure}
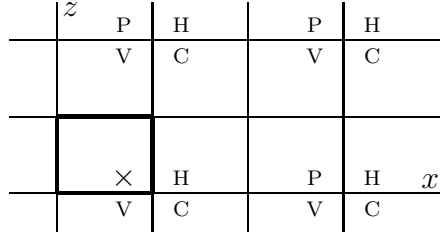

\centerline{\beginpicture
  \setcoordinatesystem units <.5in, .4in> point at 0 1
 \setplotarea x from -0.5 to 4.0, y from  -0.5 to 2.5
 \putrule from -0.5 0 to 4.0 0
  \putrule from -0.5 1 to 4.0 1
 \putrule from -0.5 2 to 4.0 2
 \putrule from 0 -0.5  to 0 2.5
 \putrule from 1 -0.5  to 1 2.5 
 \putrule from 2 -0.5  to 2 2.5 
 \putrule from 3 -0.5  to 3 2.5 
 \put{$z$} [lt] <2pt,0pt> at 0 2.5
 \put{$x$} [rb] <0pt,2pt> at 4.0 0
 \put{$\times$} at .7 .2                               
 \put{\scriptsize{P}} at 2.7 .2 
 \put{\scriptsize{P}} at .7 2.2 
  \put{\scriptsize{P}} at 2.7 2.2 
 \put{\scriptsize{H}} at 1.3 .2 
  \put{\scriptsize{H}} at 3.3 .2
   \put{\scriptsize{V}} at .7 -.2
  \put{\scriptsize{V}} at 2.7 -.2
 \put{\scriptsize{H}} at 1.3 2.2 
  \put{\scriptsize{H}} at 3.3 2.2
   \put{\scriptsize{V}} at .7 1.8
  \put{\scriptsize{V}} at 2.7 1.8
 \put{\scriptsize{C}} at 1.3 -.2 
  \put{\scriptsize{C}} at 3.3 -.2 
 \put{\scriptsize{C}} at 1.3 1.8 
 \put{\scriptsize{C}} at 3.3 1.8 
  \linethickness=1.3pt
 \putrule from 0 0 to 1 0  
  \putrule from 0 1 to 1 1  
 \putrule from 0 0 to 0 1   
 \putrule from 1 0 to  1 1 
 \endpicture
}
 \caption{Image points of a point $\times$ in a rectangle
 (cf.\ \cite{BB3,Leipzig,rect}), classified as periodic 
displacements ({\scriptsize{P}}), horizontal reflections with 
(possible) vertical periodic drift ({\scriptsize{H}}),
  vertical reflections with 
 horizontal periodic drift ({\scriptsize{V}}),
 or corner reflections ({\scriptsize{C}}),
\label{fig:images}} \end{figure}

On the plates at $x=0$ and $x=a$ each term contributes a pressure 
in the 
 $x$ direction according 
to~\eref{pj} (with $j=1$, which will be henceforth understood).
One finds that (for any~$x$)
 \be
p_{\mathrm{H}mn}=0  = p_{\mathrm{C}mn},
 \label{Cp}\ee
 \be \fl
 p_{\mathrm{P}mn}=-\,\frac{8m^2a^2}{\pi^2(t^2+y^2+4m^2a^2+4n^2b^2 
)^3} +\frac1{2\pi^2(t^2+y^2+4m^2a^2+4n^2b^2)^2} \,,
\label{Pp}\ee
 \be\fl
 p_{\mathrm{V}mn}=\frac{8m^2a^2}{\pi^2(t^2+y^2+4m^2a^2+4(z-nb)^2 
)^3} -\frac1{2\pi^2(t^2+y^2+4m^2a^2+4(z-nb)^2)^2} \,,
\label{Vp}\ee
Note that  $ p_{\mathrm{P}mn}$ and 
$p_{\mathrm{V}mn}$  are independent of~$x$.

 Similarly, the energy density due to each term can be computed 
by~\eref{rho}.  For certain purposes, however, it is convenient to 
take only one $t$ derivative before evaluating one of the sums or 
integrals in closed form. 
 In particular,  consider (at $x'=x$, $z'=z$)
   \be
T_{\mathrm{H}mn}\equiv {\partial \tbar_{\mathrm{H}mn} \over \partial t} =
 -\,\frac{t}{\pi^2(t^2+y^2+4(x-ma)^2+ 4n^2b^2)^2}\,.
 \label{TH}\ee
One has
 \begin{eqnarray}
\sum_{m=-\infty}^\infty \int_0^a T_{\mathrm{H}mn}\,dx &= 
- \sum_{m=-\infty}^\infty\int_{ma}^{(m+1)a}
 \frac{t\,dx}{\pi^2(t^2+y^2+4x^2+ 4n^2b^2)^2} 
\nonumber \\  &= -\int_{-\infty}^{\infty}
 \frac{t\,dx}{\pi^2(t^2+y^2+4x^2+ 4n^2b^2)^2} \,,
\label{EH}\end{eqnarray}
 which is independent of~$a$.
Therefore, the $\mathrm{H}$-energy in the box is unchanged 
 whenever one of the vertical walls moves, 
 which is precisely consistent with 
$\sum_m $ from~\eref{Cp}.
 Note, however, that it was essential to consider all values of~$m$ 
at once; for $m=0$, for instance, the energy in the box is not 
exactly constant because a part of the surface energy distribution 
created by one wall is pushed through the opposite wall. 

The $\mathrm{V}$-energy density is easily seen to be independent of~$x$,
  since,  in analogy to~\eref{TH},
  \be
T_{\mathrm{V}mn} 
= -\,\frac{t}{\pi^2(t^2+y^2+4m^2a^2+ 4(z-nb)^2)^2}\,.
 \label{TV}\ee
 Therefore,  the total $\mathrm{V}$-energy is equal to that density 
 times~$a$; so its 
derivative with respect to~$a$ has two terms,
  one just equal to the energy density, 
 \be
  \rho_{\mathrm{V}mn} =
  -\,\frac12\,\frac{\partial T_{\mathrm{V}mn}}{\partial t}\,,
\label{fundthmbdry} \ee
 and the other equal to 
 \be
a\, \frac{\partial}{\partial a} \rho_{\mathrm{V}mn}\,.
 \label{fundthmbulk}\ee
One calculates
 \be\fl
 \rho_{\mathrm{V}mn}
 =-\, \frac{2t^2}{\pi^2(t^2+y^2+4m^2a^2+ 4(z-nb)^2)^3}
 +\frac1{2\pi^2(t^2+y^2+4m^2a^2+ 4(z-nb)^2)^2}\,.
\label{rhoV}\ee
 If $t=0$ (the neutral cutoff), then \eref{fundthmbulk} equals the 
negative of the first term of~\eref{Vp}, and \eref{fundthmbdry}
 is the negative of the second term. 
 Thus the principle of energy balance \eref{balance} holds.
 If $t\ne0$ and $y=0$ (the ultraviolet cutoff), the first term in 
\eref{rhoV} survives and creates the pressure anomaly 
 (specifically, from the terms with $m=0$).

 A calculation similar to \eref{EH} yields the formula
 \be
 {t^2-y^2 \over 8\pi (t^2+ y^2)^2}
 \label{CE}\ee
 for the total $\mathrm{C}$-energy.  It is independent of~$a$,
 hence consistent with~\eref{Cp}.
 In dimension 2 the corresponding term vanished entirely.
 The difference is that in dimension~3 the ``corner'' is really an 
edge of a parallelepiped, whose true corners would indeed give no 
energy in a cutoff theory.

 The $\mathrm{P}$ terms are nondivergent, and of course for 
$\rho$ they give 
the standard (finite and constant) bulk Casimir energy density 
of the waveguide.

 \ack This research was supported by  National Science 
Foundation Grants 
 PHY-0554849 and PHY-0968269. We thank Lev Kaplan, Kim 
Milton, Martin Schaden, and the 
participants in the 2012 Oklahoma--Texas--Louisiana Quantum 
Vacuum Meeting for helpful discussions that raised many of the 
issues aired in \sref{implications}.

 \appendix
 \section{The pressure anomaly in dimension $d$}
 \label{append}
 When the spatial dimension is $d$ ($\ge2$), 
 the free cylinder kernel is, up to a  constant factor,
 \be
\tbar_0 \propto \left(t^2 + |\mathbf{r}-\mathbf{r}'|^2\right)
 ^{-(d-1)/2}.
\label{Tfreed} \ee
 Thus the reflected kernel generalizing \eref{Twall} is
 \be
 \tbar \propto \left(t^2 + |\mathbf{r}_\bot-\mathbf{r}_\bot'|^2
 +(z+z')^2\right)^{-(d-1)/2}.
 \label{Twalld}\ee
Therefore, following \eref{rho}--\eref{pj} and 
 \eref{Mdef}--\eref{wallp}, one finds
 \be
 \rho \propto -  M(d)^{-(d+3)/2}\left(-t^2d+\mathbf{r}_\bot{}\!^2 
+ 
4z^2\right),
 \label{wallrhod}\ee
 \be
 p_1\propto    M(d)^{-(d+3)/2}\biggl(t^2-x^2d 
+\sum_{j\ne1,d}x_j{}\!^2
  + 4z^2\biggr),
 \label{wallpd}\ee
 with $z=x_d\,$, $x=x_1\,$, and
 \be
 M(d) = t^2 + \mathbf{r}_\bot{}\!^2 + 4z^2.
 \label{Mdefd}\ee
 The missing numerical constants in \eref{wallrhod} and 
\eref{wallpd} are the same, namely \cite{LF}
 \be
 \frac12 C(d) =\frac12\pi^{-
(d+1)/2}\Gamma\left(\frac{d+1}2\right).
 \label{Tconst}\ee

 We are to choose one of the coordinates, $t$ or $x_j$ ($j\ne d$),
  to serve as 
the cutoff and to set the others to~$0$.  Call that coordinate $w$.
Then we are confronted with integrals of the type
 \be
 I_w=\int_0^\infty dz\, {-w^2d+ 4z^2
 \over (w^2+4z^2)^{(d+3)/2}}
 \label{badint}\ee
 and
\be
 I=\int_0^\infty dz\, {w^2+ 4z^2 \over (w^2+4z^2)^{(d+3)/2}} =
 \int_0^\infty dz\, {1\over (w^2+4z^2)^{(d+1)/2}}\,.
 \label{goodint}\ee
The form \eref{goodint} applies if $w$ is ``neutral'' with respect 
to the component of energy or pressure concerned;
 \eref{badint} applies if $w$ is not neutral ($w=t$ for energy 
density, $w=x$ for $p_1\,$, etc.).

  Note that 
 \be
 I_w  = I - (d+1) w^2 \int_0^\infty dz\, 
{1 \over (w^2+4z^2)^{(d+3)/2}}\,.
 \label{Irel}\ee
Because
 \[ {d\over dz}\, {z\over (w^2+4z^2)^{(d+1)/2}}=
 {(d+1)w^2 \over (w^2+4z^2)^{(d+3)/2}} - {d\over (w^2+4z^2)^{(d+1)/2}}
\,, \] 
  the second term in \eref{Irel} equals
\[ -\int_0^\infty dz\, {d\over (w^2+4z^2)^{(d+1)/2}} = -d\cdot I. 
\]
 Thus
 \be
 I_w = -(d-1)I.
 \label{Irelfinal}\ee
 When $d=3$, \eref{Irelfinal} yields the anomalies in 
\eref{papabear} and \eref{mamabear}, and \eref{goodint} yields 
the value in~\eref{papabear}.  When $d=2$, the anomaly 
reduces to a sign change, which was overlooked in the discussion 
near the top of p.~18 of~\cite{rect}.

\end{document}